\documentclass[preprintnumbers,twocolumn,secnumarabic,amssymb, nobibnotes, aps, prd]{revtex4}

\usepackage{color}
\usepackage{graphicx}
\usepackage{amsmath}
 
\begin{document}
\preprint{DCPT-14/45}
\title{Holographic Accelerated Heavy Quark-Anti-Quark Pair}

\author{Veronika E. Hubeny}
\affiliation{ Centre for Particle Theory \& Department of Mathematical Sciences,\\
Science Laboratories, South Road, Durham DH1 3LE, UK.}
\author{  Gordon W. Semenoff}
\affiliation{Department of Physics and Astronomy, University of British Columbia, \\
                     6224 Agricultural Road, Vancouver, British Columbia, Canada V6T 1Z1}
%\emailAdd{veronika.hubeny@durham.ac.uk}
%\emailAdd{gordonws@phas.ubc.ca}

\begin{abstract}
The problem of a heavy quark-anti-quark pair which have constant eternal acceleration in opposite directions in the vacuum of 
deconfined maximally supersymmetric Yang-Mills theory is studied both in perturbation theory and at strong coupling using AdS/CFT. 
Perturbation theory is summed to obtain what is conjectured to be an exact result.  
It is shown to agree with a particular prescription for computing  the disc amplitude in the string theory dual
and it yields a value $s=\sqrt{\lambda}$ for the entanglement entropy of the quark and anti-quark.  
\end{abstract}

\maketitle

The system of a quark and an anti-quark which accelerate eternally on mirror-symmetric  
trajectories has recently received some attention \cite{Caceres:2010rm}-\cite{Lewkowycz:2013laa}.  The 
quark and anti-quark are never in causal contact.  However,  together, they form a colour singlet. 
Their quantum states
are therefore highly entangled.  Moreover, they can  interact with each other by exchanging
space-like gluons and other quanta of the field theory that they are embedded in.   
How their properties would be modified by such interactions is an interesting question.    
For some quantum field theories, such as ${\mathcal N}=4$ supersymmetric Yang-Mills theory, 
this question can be addressed in both the weak coupling limit using perturbation theory and
in the strong coupling limit by studying the
string theory dual of the quark-anti-quark pair, an open string traveling on the $AdS_5\times S^5$
background.   
\begin{figure}
\begin{center}
\includegraphics[width=1.7in]{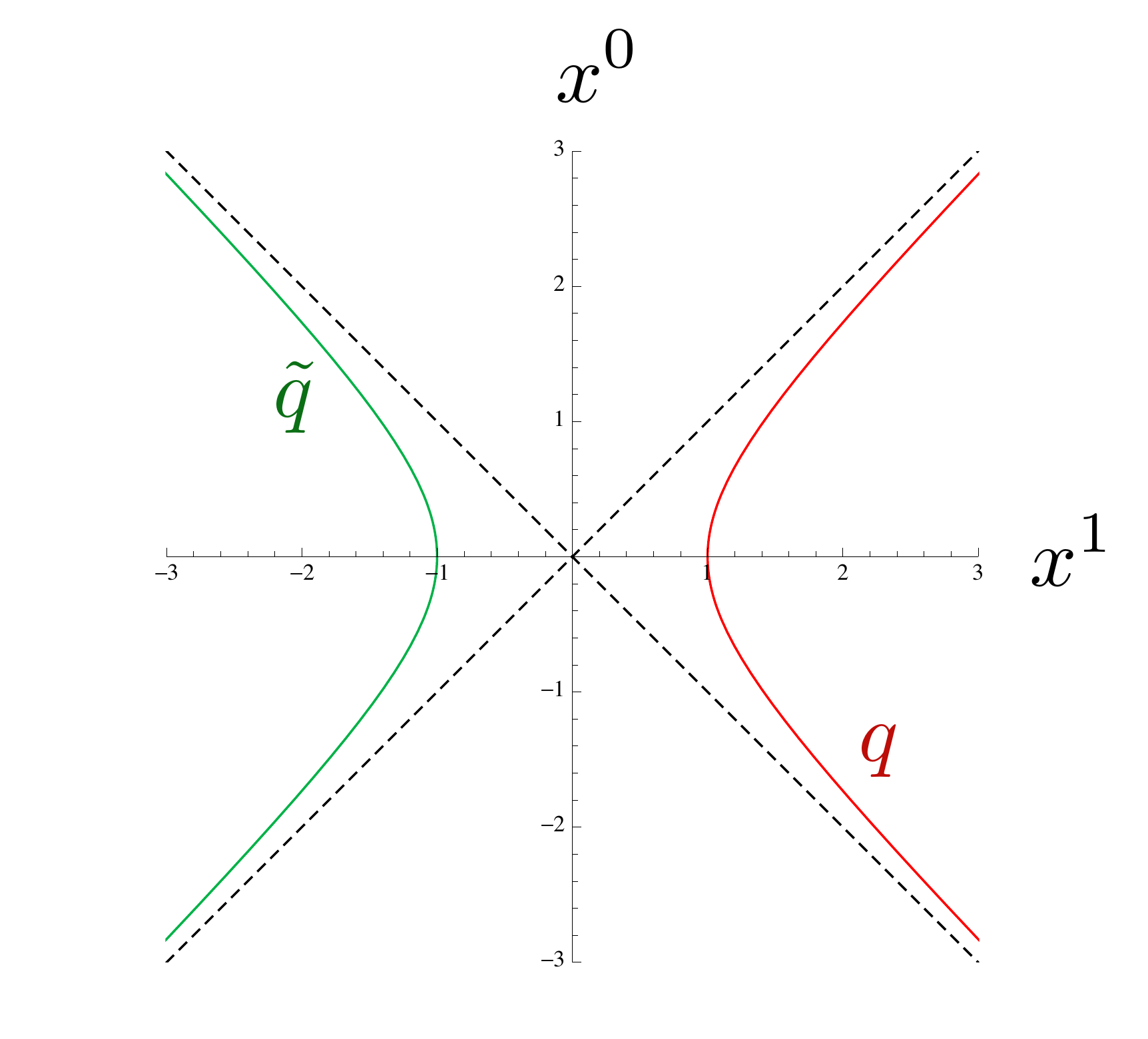}
\caption{The quark and anti-quark follow eternally accelerating trajectories on the right and left of the figure, respectively.}
\label{0}
\end{center}
\end{figure}
In this paper, we will study  an accelerating quark-anti-quark pair in ${\cal N}=4$ Yang-Mills theory.  
To be concrete, we assume that the acceleration is
generated by a constant electric field $E$.   The quark and anti-quark are scalar components of the
 massive W-boson supermultiplet which is created when, in ${\mathcal N}=4$ supersymmetric Yang Mills
theory,  $U(N+1)$ gauge symmetry is spontaneously broken
to $U(N)\times U(1)$.   The  electric field is
in the unbroken $U(1)$. 
We will consider the quantum amplitude for a process where we inject the quark and anti-quark into
the system at early times and at very high velocities on a collision course.   The electric field decelerates
both the quark and anti-quark.   They stop and turn around before they collide and then accelerate back, in 
opposite directions to spatial infinity.   We will take the heavy quark limit where this semi-classical description of quark
propagation is valid.   In particular, this limit should suppress competing processes such as Schwinger pair production 
by the electric field and the Drell-Yan process where the quark and anti-quark annihilate to form a space-like gluon which
then decays to a quark-anti-quark pair. 
 We will eventually take the large $N$ limit where internal loops of the W-boson and emission of
  bremsstrahlung are suppressed.

The trajectories which solve the classical equation of motion 
of the quark and the anti-quark are 
\begin{align}\label{rindlertrajectories}
x^\mu(\tau) =\tfrac{M}{E}\left( \sinh\tfrac{E}{M}\tau, 
\cosh\tfrac{E}{M}\tau ,0,0\right)  ~,
\\  \label{rindlertrajectories1}
\tilde x^\mu(\tilde \tau)=
\tfrac{M}{E}\left( \sinh\tfrac{E}{M}\tilde \tau, -\cosh\tfrac{E}{M}\tilde\tau ,0, 0\right) ~,
\end{align} 
respectively and are  depicted in figure \ref{0}. 
 Here,  $\tau,\tilde\tau$ are the proper times and $M$
is the quark mass.  These trajectories are tuned so that the distance of closest approach
is given by $2\frac{M}{E}$.  The amplitude can be extracted from the 4-point function
 \begin{align} 
& g(x_f,\tilde x_f;x_i,\tilde x_i)= \nonumber  \\
 &\int_0^\infty\frac{ dT}{T} \int [dx^\mu] e^{iS[x,T]}
 \int_0^\infty\frac{ d\tilde T}{\tilde T}  \int [d\tilde x^\mu] e^{ i\tilde S[\tilde x,\tilde T] }~W[x,\tilde x]
 \nonumber\\
 &\approx ~\exp\left(~-i{\mathcal E}\tau_P~\right)
 \label{4point}
\end{align} 
where we use the world-line path integral representation of the quark and anti-quark propagators, with
actions 
 \begin{align}\label{worldlineaction}
&S
%[x,T]
=\int_0^1d\sigma\left[\frac{(\dot x^\mu(\sigma))^2}{4T}-M^2T+\tfrac{E}{2} (x^1\dot x^0
-x^0\dot x^1)\right]
\\
&\tilde S
%[\tilde x,\tilde T]
=\int_0^1d\sigma\left[\frac{(\dot {\tilde x}^\mu(\sigma))^2}{4\tilde T}-M^2\tilde T-\tfrac{E}{2} (
\tilde x^1\dot{\tilde  x}^0
-\tilde x^0\dot{\tilde  x}^1)\right]
\label{tildeworldlineaction}
\end{align}
and boundary conditions $x^\mu(\sigma=1)=x^\mu_f$ and $x^\mu(\sigma)=x^\mu_i$. 
\begin{figure}
\begin{center}
\includegraphics[width=1.5in]{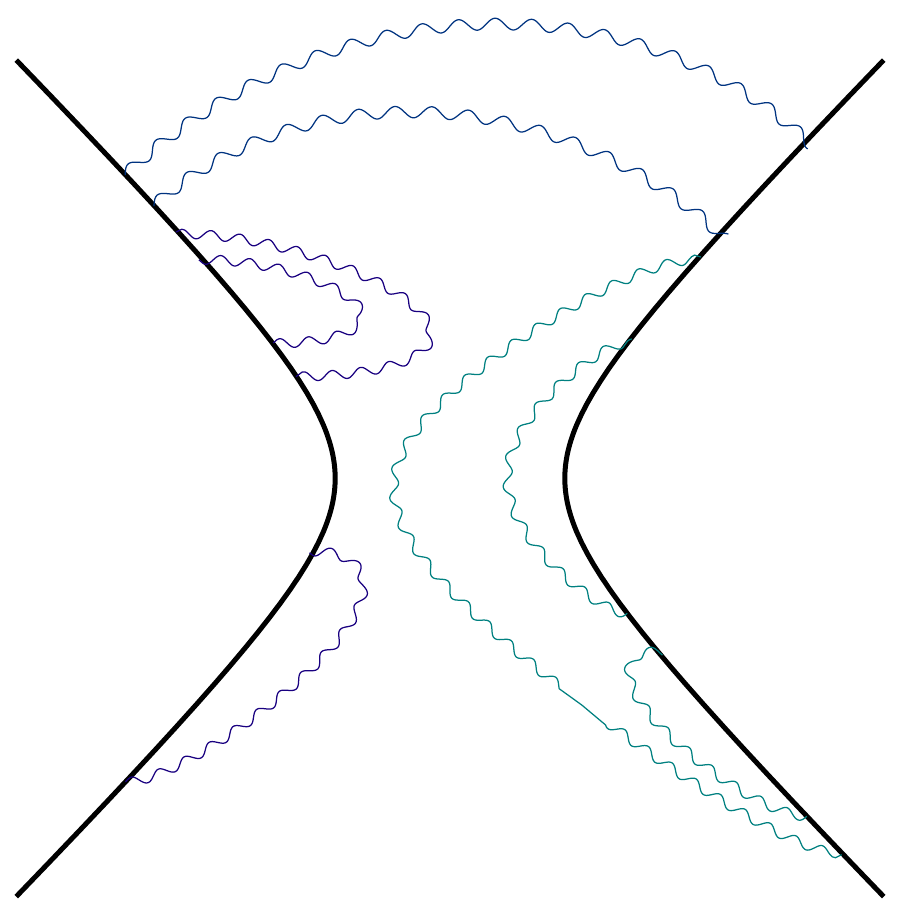}
\caption{The four point function is dominated by the amplitude where the quark and anti-quark follow the solid lines
and exchange massless ${\cal N}=4$ quanta along the wiggly lines.  Only planar ladder  diagrams are depicted here.}
\label{twoquarks}
\end{center}
\end{figure}
Interactions with the remaining massless U(N) adjoint representation fields of ${\cal N}=4$ theory
are taken into account by inserting the Wilson loop for the quark and anti-quark, 
\begin{align}\label{wilsonloop}W[x,\tilde x] = \left<0\right|{\rm Tr}~U[x]\tilde U [\tilde x]
\left|0\right>
\end{align}
where
\begin{align} 
&U[x]={\mathcal T}e^{i\int_{-\infty}^\infty d\tau
\left[ A_\mu(x(\tau)) \dot x^\mu(\tau)+\sqrt{-\dot x(\tau)^2}\phi^1(x(\tau))\right]}
\label{u}
\\ \label{wilsonloop2}
&\tilde U[\tilde x]=\left(
{\mathcal T}e^{i\int_{-\infty}^\infty d\tau\left[ A_\mu(\tilde x(\tau)) 
\dot{\tilde x}^\mu(\tau)-\sqrt{-\dot {\tilde x}(\tau)^2}\phi^1(\tilde x(\tau))\right]}
\right)^\dagger
\end{align}
Our aim is to compute the phase and extract ${\mathcal E}$ defined in (\ref{4point}) in the heavy quark limit, 
the large $N$ planar limit and the large $M\tau_P$ limit.

Consider propagator for the free quark or antiquark, 
$\int_0^\infty\frac{ dT}{T} \int [dx^\mu] \exp\left(iS[x,T]\right)$ in the semi-classical large $M$ limit.
The classical  equations of motion for the action  (\ref{worldlineaction}) or (\ref{tildeworldlineaction})
are solved by
\begin{align}\label{classical}
x_0^\mu\left(\tau_P(\sigma-1/2)\right)~,~\tilde x_0^\mu\left(\tau_P(\sigma-1/2)\right) 
~,~T_0=\frac{\tau_P}{2M}=\tilde T_0 
\end{align}
where $(x_0(\tau),\tilde x_0(\tilde \tau))$ are  the trajectories (\ref{rindlertrajectories},\ref{rindlertrajectories1}).  
The classical actions evaluated on the solution
are $S[x_0,T_0]=\tilde S[\tilde x_0,\tilde T_0]=\left[M -\frac{M}{2}\right]\tau_P$.  
The first term, $M\tau_P$, is the rest mass
times the proper time.  The second term, $\frac{M}{2}\tau_P$, is from the interaction with the electric field. The
quark propagator is then $\int_0^\infty\frac{ dT}{T} \int [dx^\mu] \exp\left(iS[x,T]\right)\sim\exp\left(-i\left[M-\tfrac{M}{2}\right]\tau_P\right)$.
We are only interested in the asymptotic behaviour of the phase at large $\tau_P$.  Corrections to the coefficient of $\tau_P$ in the exponent
are suppressed by powers of  $\tfrac{E}{M^2}$ which we assume to be small.  

Now, consider the coupling to the remaining massless fields of the ${\mathcal N}=4$ theory.  To analyze  the
integrals in (\ref{4point}) semi-classically we should take into account
the contributions of the Wilson loop which would contribute a force, $-i{\delta\ln W[x,\tilde x]}/{\delta x^\mu(\tau)}$,
to the equation of motion for the quark and similar for the anti-quark.  
However, since,  by symmetry,
$$
\left. \frac{\delta\ln W[x,\tilde x]}{\delta x^\mu(\tau)}\right|_{x,\tilde x=x_0,\tilde x_0}=0~,~
\left. \frac{\delta\ln W[x,\tilde x]}{\delta\tilde  x^\mu(\tau)}\right|_{x,\tilde x=x_0,\tilde x_0}=0~,
$$
the classical equations of motion which govern the semiclassical limit of the four-point function are still solved by the same classical
trajectories (\ref{classical}) as in the absence of the Wilson loop.   The semiclassical limit is then given by
\begin{align} 
& g(x_f,\tilde x_f;x_i,\tilde x_i)\approx  e^{ -2i\left[M-\tfrac{M}{2}\right]\tau_P}~W[x_0,\tilde x_0]
 \label{4pointapprox}
\end{align} 
and, estimating the leading order reduces to evaluating the Wilson loop on the trajectories (\ref{rindlertrajectories},\ref{rindlertrajectories1}). 
To compute the Wilson loop, we expand  the time ordered products in powers of the exponents in (\ref{u})
and (\ref{wilsonloop2}) and 
evaluate the resulting correlation functions using the standard Feynman-Dyson technique.
For details of conventions and notation, 
we refer the reader to reference \cite{Erickson:2000af} where a similar
computation is done for the Euclidean circle Wilson loop. 
The free scalar field  and the free vector field propagators in the Feynman gauge are 
$
\left<{\cal T} \phi^{ab}(x)\phi^{cd}(y)\right>_0 = \frac{g^2\delta^{ad}\delta^{bc}}{2}g(x-y)
$ and 
$
\left<{\cal T} A_\mu^{ab}(x)A_\nu^{cd}(y)\right>_0 = \frac{g^2\delta^{ad}\delta^{bc}g_{\mu\nu}}{2}g(x-y)
$ with 
$g(x)=1/[4\pi^2 x^2+i\epsilon]$.
The essential observation is that, if we sum the contribution of vector and scalar propagators between any two points on
the trajectories, the effective propagator, for example, $\tfrac{\sqrt{-\dot x(\tau_1)^2}\sqrt{-\dot x(\tau_2)^2}+\dot x_\mu(\tau_1)\dot x^\mu(\tau_2)}{8\pi^2\left[ x(\tau_1)
-x(\tau_2)\right]^2+i\epsilon}
=\frac{E^2}{16\pi^2M^2}$, is a constant.   
Then, since the effective propagators are constants, 
one can sum rainbow ladder diagrams of the type depicted in 
figure \ref{twoquarks} by simply solving the
algebraic problem of contracting Lie algebra indices.     
As for the case of the Euclidean circle studied in reference \cite{Erickson:2000af}, 
this problem is  summarized 
by a Gaussian matrix model.  Moreover, the matrix model integral can
be solved exactly, for any value of $N$ and $\lambda$.   The result 
for the Euclidean circle at large $N$ was given in reference \cite{Erickson:2000af} and for 
any $N$ in \cite{Drukker:2000rr} and can  be modified for the   
 Lorentzian case to get 
\begin{figure}
\begin{center}
\includegraphics[width=1.5in]{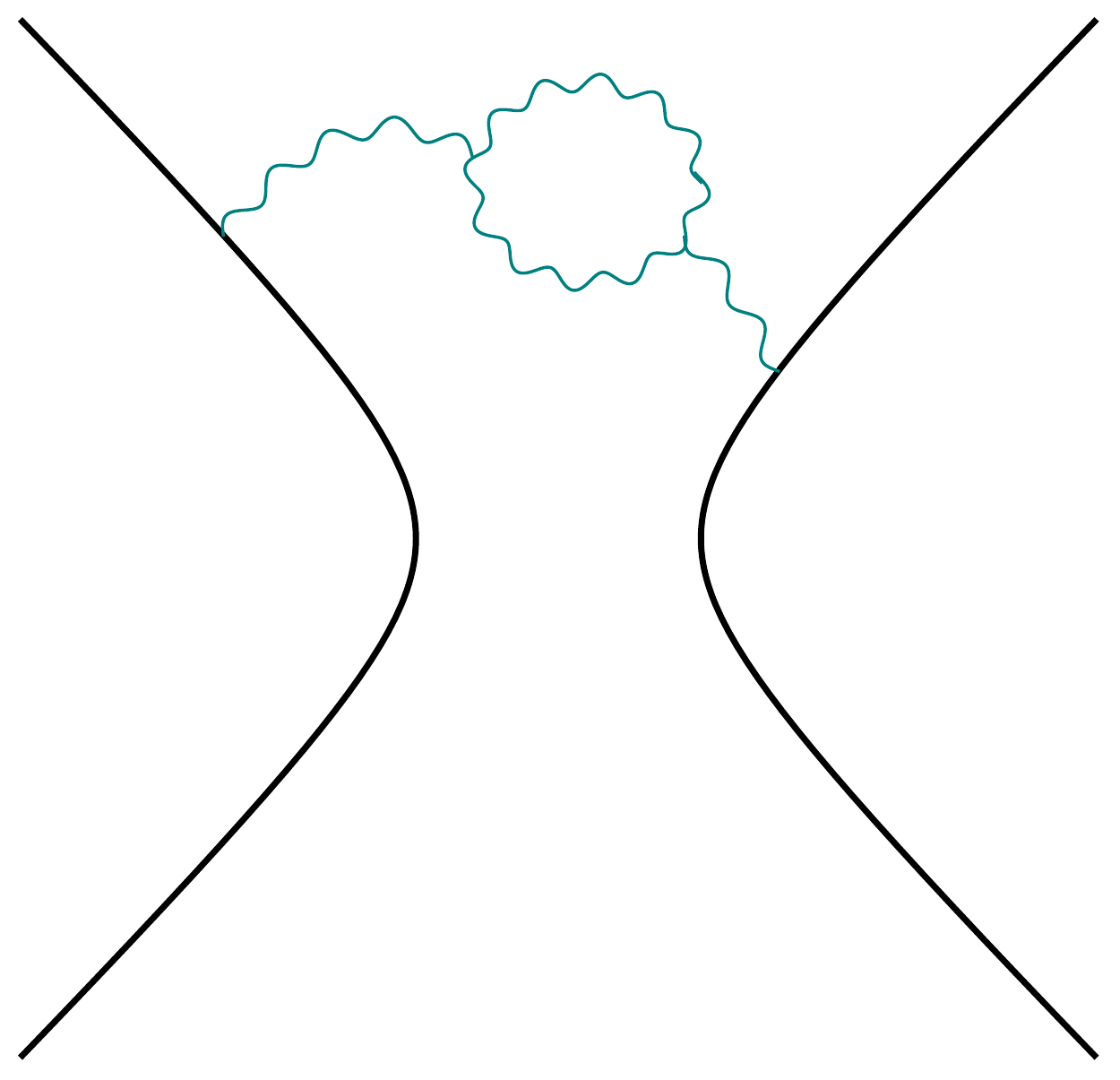}
\caption{Leading contribution with internal vertices: Feynman diagram with a self-energy 
correction.}
\label{corr1}
\end{center}
\end{figure}
\begin{figure}
\begin{center}
\includegraphics[width=1.5in]{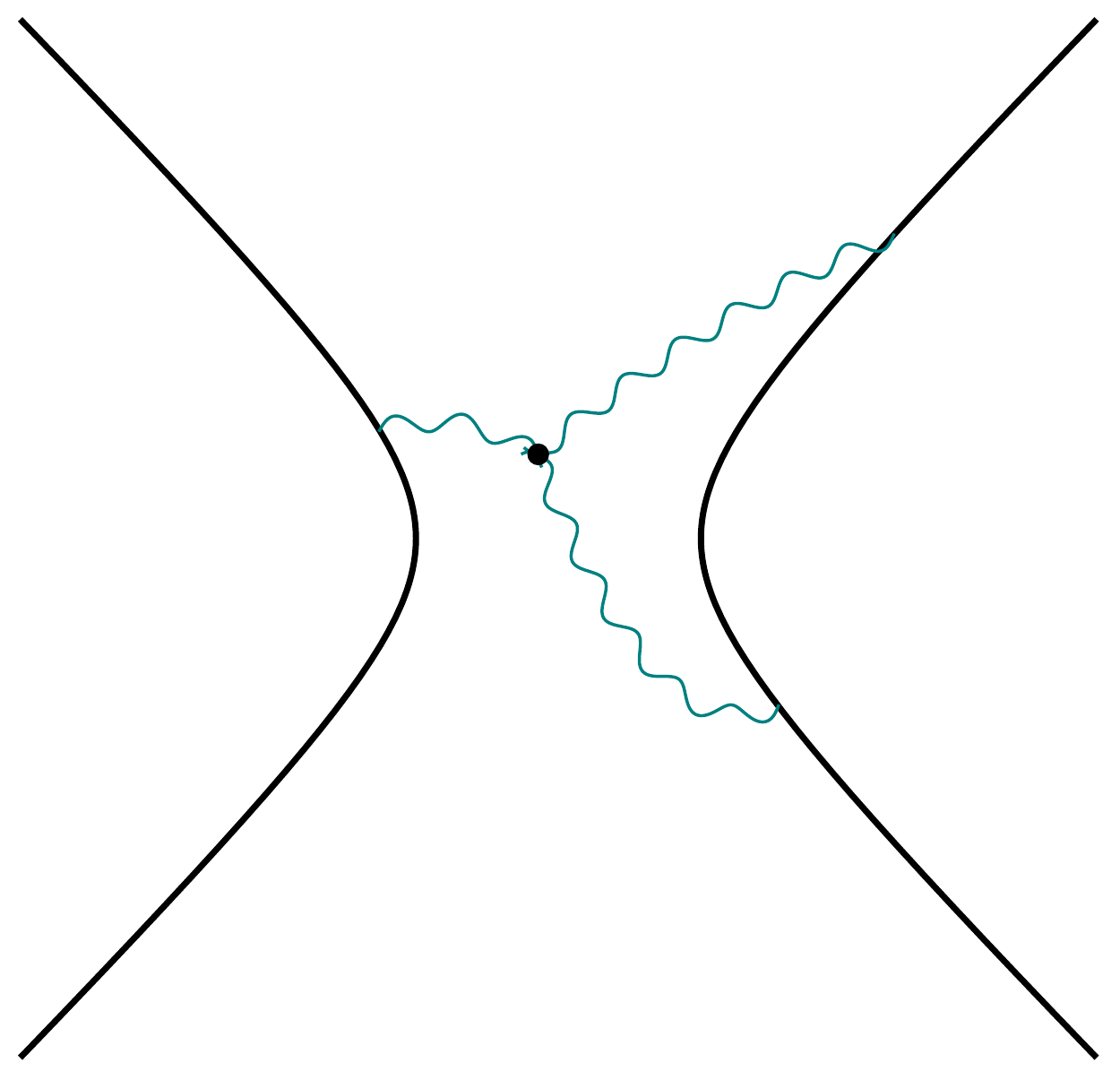}
\caption{Leading contribution with internal vertices:  Feynman diagram with one internal vertex.  }
\label{corr2}
\end{center}
\end{figure}
\begin{align}\label{finiteNwilsonloop}
&W[x_0,\tilde x_0] =NL^1_{N-1}\left[ \frac{\varepsilon^2}{N}
(M\tau_P)^2\right]~e^{ -\frac{
\varepsilon^2}{2N}(M\tau_P)^2}
\end{align}
where $L^m_n(x)= 
\frac{1}{n!}e^x x^{-m}\left(\frac{d}{dx}\right)^n \left(x^{m+n}e^{-x}\right)$ 
is the Laguerre polynomial and 
\begin{align}
\varepsilon=\frac{\sqrt{\lambda}}{2\pi}\frac{E}{M^2}
\label{defnvarepsilon}
\end{align}
This sum over ladder diagrams has left out diagrams with internal vertices.  Their leading orders
are depicted
in figures \ref{corr1} and \ref{corr2}.  In the case of the circle Wilson loop discussed in \cite{Erickson:2000af},
the contributions
of these diagrams canceled exactly, leading to the conjecture, later proved using localization \cite{Pestun:2007rz}, 
that all diagrams with internal vertices cancel and that the sum of ladders
is  exact. The Wilson loop (\ref{wilsonloop}) with trajectories (\ref{rindlertrajectories},\ref{rindlertrajectories1}) is supersymmetric in that it commutes with half of the supercharges. 
It is straightforward to see by explicit calculation that, as $\tau_P\to\infty$,  
the leading corrections cancel in the present Lorentzian case also.
 This leads us to the conjecture that,  
for large $\tau_P$, (\ref{finiteNwilsonloop}) approaches an exact formula.  We will proceed with the assumption that
it is exact, with emphasis that, at this point, it is an assumption.

Now, we observe that (\ref{finiteNwilsonloop}) has Gaussian damping with proper time and its contribution to the amplitude 
at large $\tau_P$ vanishes. 
This can be attributed to the fact that accelerating quarks
 emit  bremsstrahlung and the amplitude for finding only a quark-anti-quark pair in the final state is vanishingly
 small.  It is also easy to see that emission of  bremsstrahlung requires nonplanar Feynman diagrams which are suppressed in the large $N$ limit. This  suggests that the proper asymptotic behaviour should be restored if we
 first project onto planar diagrams and then take $\tau_P$ large.  
Indeed, by looking at the large $N$ limit of (\ref{finiteNwilsonloop}), we see that the Gaussian damping goes away,  
 \begin{align}\label{largeNlimit}
&\lim_{N\to\infty}
W [x_0,\tilde x_0]
=\frac{N}{\varepsilon M\tau_P } J_1\left(2\varepsilon M\tau_P
\right) 
\\
\label{asymptoticinTwilsonloop}
&\lim_{N,\tau_P \to\infty}  W[x_0,\tilde x_0]= N 
\frac{-i }
{\sqrt{ 4\pi }\left(\varepsilon M\tau_P
\right)^{\frac{3}{2}} }
~e^{i~2\varepsilon M\tau_P} 
%+{\cal O}(\tfrac{N}{\tau_P^3})+ {\cal O}(\tfrac{1}{N})
\end{align}
 Here,   
$J_\alpha(x)$ are Bessel
functions of the first kind and
$J_a(x)\sim \sqrt{\frac{2}{\pi x}}\cos\left(x-\frac{\pi}{2}a\right)$ at large agrument.  The cosine is a sum of two exponentials.  
The large $M\tau_P$ limit must
be defined with an $i\epsilon$ prescription  that picks out the appropriate
exponential.
 The resulting phase, when combined with (\ref{4pointapprox}), gives the result
for  the energy of the quark-anti-quark pair, 
\begin{align}
&{\cal E}\approx 2\left[M- \varepsilon M -\frac{M}{2}\right]\label{ymresult1}
\end{align}
The pre-factor in (\ref{asymptoticinTwilsonloop}) is also interesting as it summarizes the
effect of fluctuations corresponding to those of the disc geometry in the string theory dual. 
However, we have not computed the contribution of fluctuations of the quark and anti-quark paths, which would contribute at the same
order, and are dual to fluctuations of the disc boundary in the 
string theory.  We leave this as a project for future work. 
 
When $\sqrt{\lambda}$ is large
so that $\varepsilon \gg E/M^2$, the correction in (\ref{ymresult1}) would be 
larger than the order $E/M^2$ terms which would arise from taking into account 
fluctuations of the trajectory 
about the classical limit and which we have ignored.  However, we cannot let $\varepsilon$  be of order one, as the amplitude for Schwinger
pair production, which also occurs in a constant external electric field, 
is modified at strong coupling \cite{Semenoff:2011ng}.  
There is a critical electric field
where the tunnelling barrier to pair production vanishes and it
loses its usual exponential suppression.  That critical electric field is at  $\varepsilon=1$. 
If we want to avoid this regime, we must also take $\varepsilon \ll 1$.
This restricts the region of validity of our computation and, simultaneously, the regime where
the correction from the Wilson loop is significant, to
\begin{align}\label{regime}
\frac{E}{M^2} \ll \varepsilon=\frac{\sqrt{\lambda}}{2\pi}\frac{E}{M^2} \ll 1
\end{align} 
This is the strong coupling regime of Yang-Mills theory.

The AdS/CFT dual of ${\mathcal N}=4$ Yang-Mills theory is  
IIB string theory on the $AdS_5\times S^5$ background,  
\begin{equation}
ds^2 = \frac{\sqrt{\lambda}\alpha'}{u^2} \, \left( \eta_{\mu\nu} \, dx^{\mu}\, dx^{\nu} + du^2 \right)
\label{AdSPoincmet}
\end{equation}	
 The AdS boundary is located at $u=0$ and the Poincare horizon at 
$u=\infty$.  The quark and anti-quark are oppositely oriented open strings which stretch from the 
Poincare horizon of $AdS_5\times S^5$ to a probe D3 brane that is suspended parallel to the horizon
at radial coordinate $u=u_M$.  The quark mass  $M$ is the energy of a static string suspended between the probe and the Poincare horizon.
This determines the radius at which the probe is suspended,  $ u_M=\frac{\sqrt{\lambda}}{2\pi M}$.    
 The probe brane has an internal constant U(1) electric field $E$, 
 where $E$ is assumed to be small enough that 
it does not back-react on the probe brane embedding. (This is the condition $\varepsilon \ll 1$.).
We need to compute the disc amplitude, where the boundary of the disc lies on the
probe brane and where the asymptotic states are the in-coming and out-going 
quark-anti-quark pair. 
In the limit of large ${\lambda}$, the open string sigma model is  
semiclassical and it suffices to study classical solutions of the Nambu action, 
\begin{align}\label{nambuaction}
S_N=\tfrac{-1}{2\pi\alpha'}\int d^2\sigma\sqrt{-\det g}+\tfrac{E}{2}\oint  d\tau (x^1\dot x^0
-x^0\dot x^1)
\end{align}
with $g$   the world-sheet metric.  
  The last term in  (\ref{nambuaction})  is  
the line integral on the world-sheet boundary of the U(1) gauge field corresponding to  a constant
electric field.  Then, a classical solution is a world sheet which is itself a part of 
$AdS_2$, the locus of 
\begin{align}\label{locus}
u^2+(x^1)^2-(x^0)^2=\frac{M^2}{E^2}  
\end{align}
This world-sheet intersects the probe brane at the quark and anti-quark trajectories 
(\ref{rindlertrajectories},\ref{rindlertrajectories1}). 
There are two world-sheet event horizons which are located at $u_E=\frac{M}{E}$.  
The metric near the quark trajectory is
\begin{equation}
ds_{g}^2 =\sqrt{\lambda}\alpha'\left[ - \left( \frac{1}{u^2} - \frac{E^2}{M^2} \right) \,  d\tau^2
- \frac{2}{u^2} \, d\tau \, du  \right]
\nonumber
\end{equation}	
One can check that this agrees with the general solution for the extremum of the Nambu action with arbitrary
curve on the AdS$_5$ boundary which was found by Mikhailov \cite{Mikhailov:2003er}. As Mikhailov observed,
the determinant of the world-sheet metric does not depend on the trajectory, and in this particular case it does not
depend on the electric field.  An integration to find the proper area of the full AdS$_2$ world sheet
 yields an on-shell action which contains only the quark mass (plus interaction with electric field), 
$
S_N =-2\left[M-\frac{M}{2}\right]\tau_P
$, 
  which disagrees with the Yang-Mills computation since the 
  energy shift of the quark is absent.  (The factor of 2  is from integration over two identical regions.)
What the Yang-Mills computation suggests is, rather, that we should use the proper area between the probe brane and the event horizons 
\begin{align}
S_N = 2 \left[-\frac{\sqrt{\lambda}}{2\pi}\int_{-\tfrac{\tau_P}{2}}^{\tfrac{\tau_p}{2}}d\tau\int_{u_M}^{u_E} \frac{du}{u^2} +\frac{M}{2}\tau_P\right]  \nonumber \\
  =-2\left[M-\varepsilon M-\frac{M}{2}\right]\tau_P
  \end{align}
  We see that, in this case, the semiclassical disc amplitude, $\exp\left(iS_N\right)=
  \exp\left(-2i\left[M-\varepsilon M-\frac{M}{2}\right]\tau_P\right)$,  matches the Yang-Mills result (\ref{ymresult1}) exactly.
  Although this certainly deserves further study, on the face of it, the Yang-Mills computation favours a picture of the
  world-sheet advocated in reference \cite{Mikhailov:2003er} and elaborated  in \cite{Garcia:2012gw} where the world sheet, rather than being the maximally extended AdS$_2$ wormhole which is the full locus of (\ref{locus}), is 
  identical to that surface only between the event horizons and
  the probe brane. %  {\footnote This is quite remarkable, since as explained in \cite{hs}, for a general smooth string worldsheet, 
  % the location of worldsheet horizon is only determined globally, i.e.\ it requires knowing the full quark trajectory until $\tau_P = \infty$.}
     Between the event and Poincare horizons, it is replaced by the null surface 
  which has vanishing proper area. The reasoning is that, with
  causal propagation of influences on the world-sheet, the quark acceleration has no effect on the world-sheet beyond
  the horizon.  
  %The region between  horizons thus remains as it was in the initial state
  %where the quark and anti-quark were moving at a speed approaching that of light. 

  We note that the Unruh temperature of the accelerated reference frames is $T_U=\frac{E}{2\pi M}$
    and 
  $\frac{1}{2}{\mathcal E}= M-\frac{M}{2}-\sqrt{\lambda}~T_U$ can be interpreted as the free energy
  of the quark  in its rest frame.  In the limit, $T_U<< M$, which we have considered,  most of the thermal effects due to the accelerated reference frame
  are ignored.  The small correction that we see is due to interactions and it is significant only at strong coupling. The entropy which we
deduce from this free energy is $s=\sqrt{\lambda}$.   
  This entropy must be  entirely due to quantum entanglement of the accelerated quark with degrees of freedom behind its horizon.  
  Why this value differs from other estimates of this entanglement entropy \cite{Jensen:2013ora}-\cite{Lewkowycz:2013laa} 
  is an
  interesting question.
  
Finally, there is an obvious question: what about a single quark, without the anti-quark?  In the
Yang-Mills calculation, there is a big difference with the quark-anti-quark pair, 
the cancellation of contributions from the diagrams in figures \ref{corr1} and \ref{corr2} does not occur for
a single quark and we are not able to do a Yang-Mills calculation of its energy shift beyond order $\lambda^2$.   
It would be interesting to see
if this difference is reflected in the string theory dual.
 We will revisit this issue in a companion paper\cite{hs}.

 \begin{acknowledgments}
G.W.S thanks NSERC of Canada for support.
VH was supported in part by the Ambrose Monell foundation, and by the STFC Consolidated Grant ST/J000426/1.
\end{acknowledgments}

\end{document}